\documentclass[10pt,conference]{IEEEtran}
\IEEEoverridecommandlockouts
\usepackage{enumitem}
\usepackage{cite}
\usepackage{amsmath,amssymb,amsfonts}
\usepackage{algorithmic}
\usepackage{caption}
\usepackage{graphicx}
\usepackage{textcomp}
\usepackage{xspace}
\usepackage{algorithmic}
\usepackage[ruled,vlined]{algorithm2e}
\usepackage{listings}
\usepackage{hyperref}
\usepackage[table,xcdraw]{xcolor}
\usepackage{makecell}
\usepackage{booktabs,multicol,multirow,tabularx,ragged2e,dcolumn}
\usepackage[normalem]{ulem}
\useunder{\uline}{\ul}{}
\definecolor{calmGrey}{RGB}{229, 229, 229} 

\newcommand{\flaker}{\textsc{Flaker}\@\xspace}
\newcommand{\etal}{\emph{et al}.\@\xspace}
\newcommand*{\eg}{\emph{e.g.,}\@\xspace}
\newcommand*{\ie}{\emph{i.e.,}\@\xspace}
\newcommand*{\aka}{\emph{a.k.a.}\@\xspace}
\def\BibTeX{{\rm B\kern-.05em{\sc i\kern-.025em b}\kern-.08em
    T\kern-.1667em\lower.7ex\hbox{E}\kern-.125emX}}

\definecolor{javared}{rgb}{0.6,0,0} 
\definecolor{javagreen}{RGB}{27,169,89} 
\definecolor{javapurple}{rgb}{0.5,0,0.35} 
\definecolor{javadocblue}{rgb}{0.25,0.35,0.75} 

\begin{document}

\title{On the Use of Mutation in Injecting Test Order-Dependency\thanks{\textbf{Accepted at the MSR 2021 Registered Reports Track.}}}



\newcommand{\linebreakand}{%
  \end{@IEEEauthorhalign}
  \hfill\mbox{}\par
  \mbox{}\hfill\begin{@IEEEauthorhalign}
}
\makeatother
\author{\IEEEauthorblockN{Sarra Habchi}
\IEEEauthorblockA{University of Luxembourg \\
sarra.habchi@uni.lu}

\and
\IEEEauthorblockN{Maxime Cordy}
\IEEEauthorblockA{University of Luxembourg \\
maxime.cordy@uni.lu}
\and
\IEEEauthorblockN{Mike Papadakis}
\IEEEauthorblockA{University of Luxembourg \\
mike.papadakis@uni.lu}
\and
\IEEEauthorblockN{Yves Le Traon}
\IEEEauthorblockA{University of Luxembourg \\
yves.letraon@uni.lu}

}
\maketitle

\begin{abstract}
\textit{Background:} Test flakiness is identified as a major issue that compromises the regression testing process of complex software systems.
Flaky tests manifest non-deterministic behaviour, send confusing signals to developers, and break their trust in test suites.
Both industrial reports and research studies highlighted the negative impact of flakiness on software quality and developers' productivity.
While researchers strive to devise solutions that could help developers addressing test flakiness, the elaboration and assessment of these solutions are hindered by the lack of datasets large enough to leverage learning techniques. 
\textit{Aim:} To address this lack, we conduct an exploratory study that investigates a new mean for producing datasets of flaky tests.
\textit{Method:} We propose an approach that relies on program mutation to inject flakiness in software tests. 
In particular, we plan to delete \textit{helper} statements from tests to make their outcomes order-dependent, \textit{i.e.,}  pass in certain running orders but fail in other orders. 
We intend to apply our mutation-based approach to a set of 14 Java projects to assess the effectiveness of test mutation in injecting order-dependency and generate a new dataset that could be used to study flakiness.

\end{abstract}


\section{Introduction}

Regression testing is a critical activity of modern software development. 
It enables simultaneous code development by validating program changes, \ie providing confidence that these changes do not break existing functionality. 
Nonetheless, this process can be hindered whenever non-deterministic behaviours are exhibited. In particular, flaky tests, which pass and fail for the same code version, send false alerts to developers and break their trust in the test suite.
Developers waste considerable amounts of time investigating flaky test failures and this can slow down the continuous integration~\cite{Eck2019}.
Besides, oft-repeated flaky failures may lead developers to disregard real test failures or even dispels them from writing tests. 
For these reasons, flakiness is widely discussed among software practitioners~\cite{FlakinessSpotify,FlakinessGoogle,Micco2017,mozillaIntermittent}.

Several reports have shown that developers observe recurrently flaky tests and they struggle to address them.
This struggle stems from the non-deterministic nature of flaky tests and the lack of suitable tools to cope with them.
Specifically, developers are ill-equipped to distinguish flaky failures from real faults.
Thus, they tend to rerun tests several times looking for persistent failures or passes.
However, the rerun strategy is costly both computation and time-wise and its efficiency at identifying and reproducing flaky tests is questionable~\cite{Lam2020}.

Responding to this need, researchers have started putting increasing amounts of efforts in understanding flaky tests and forming efficient techniques to detect them.
Notably, the DeFlaker tool was proposed to detect flaky tests based on historical regression test executions~\cite{Bell2018}. 
Also, Lam \etal~\cite{Lam2019iDFlakies} proposed the iDFlakies tool that targets a different type of flakiness, tests for which the outcome changes depending on their order in the suite, \aka order-dependent tests. Furthermore, researchers envision the usage of machine learning to address flakiness.
Indeed, learning techniques are already used for predicting flaky tests statically~\cite{King2018,Pinto2020,Bertolino2020}.
Moreover, machine learning could assist in detecting, reproducing, and fixing flaky tests.
Nevertheless, these applications are limited by the lack of large sets of flaky tests that could serve the purposes of learning and evaluating.
Hence, there is a strong need for novel, large and diverse datasets of flaky tests.

Empirical studies rely on flakiness detection tools to build datasets from open source projects~\cite{Bell2018,Lam2019iDFlakies}.
For instance, a dataset was built using the DeFlaker tool, and it remains, to the best of our knowledge, the largest set with  1,874 flaky tests~\cite{Bell2018}.
Yet, later studies showed that this set is hardly reproducible, which makes it difficult to use for learning fixes or root causes~\cite{Pinto2020}. Nevertheless, it is hard to collect large, reproducible datasets from open source projects since flakiness appears in rapidly evolved (and tested) development environments. An additional root cause of flakiness regards the infrastructure used, which pose additional questions on the validity, responsibility and feasibility of the developed methods and their results. 

In this paper, we present an exploratory study that aims to address the lack of test flakiness datasets, in general, and order-dependent tests in particular, from a different perspective.
Instead of trying to mine test flakiness from projects, we propose \flaker, an approach that injects order-dependency flakiness in software tests. \flaker relies on program mutation to make tests order-dependent while it does not influence the test results. 
Our focus on order-dependent tests is motivated by the well-founded evidence about the sources and fixes of this type of flaky tests. 

In particular, previous studies showed that order-dependency manifests through recurrent patterns of interactions between tests, such as \textit{polluters}, \textit{victims}, and \textit{brittles}~\cite{Gyori2015,Lam2019iDFlakies}. Similarly, order-dependent tests can also be fixed thanks to \textit{helper} statements that appear in \textit{cleaners} and \textit{state-setters}~\cite{Shi2019iFix}.
These \textit{helpers} set or reset the shared state to lay the ground for a successful and stable test execution.
The objective of our approach is to remove these \textit{helpers} and make stable tests order-dependent.
For this, we rely on program mutation, which is well-suited and widely used for these purposes.

We intend to evaluate \flaker on 14 open-source Java projects that we collect from a public dataset.
The objective of this evaluation is to answer the following research questions:
\begin{itemize}[wide=10pt,noitemsep,topsep=0pt]
    \item \textbf{\textsc{RQ1}}: How effective is mutation in injecting order-dependent tests?\\
    \textbf{Objective:} We aim to assess the suitability of test mutation for generating datasets of order-dependent tests.\\
    \textbf{Evaluation:} We compare the number of order-dependent tests generated by mutation to the numbers of order-dependent tests that could be mined using detection tools like iDFlakies.      
    \item \textbf{\textsc{RQ2}}: What are the characteristics of injected order-dependent tests?\\
    \textbf{Objective:} We aim to characterise the injected order-dependent tests.\\
    \textbf{Evaluation:} Following the terminology of previous research~\cite{Shi2019iFix}, we classify the generated tests in terms of \textit{victims}, and \textit{brittles}.
    We also investigate the nature of failures caused by the mutants, \eg Assertion errors or other exceptions. 
    \item \textbf{\textsc{RQ3}}: What projects are more receptive to order-dependency injection?\\
    \textbf{Objective:} We aim to identify and characterise testing systems that are prone to order-dependency injection.\\
    \textbf{Evaluation:} We explore test suite properties that could explain the proneness to test order-dependency, \eg shared states, fixture usages, etc.
\end{itemize}



\section{Background}
\label{sec:background}

Order-dependent tests are a category of flaky tests that manifests a non-deterministic behaviour because of the execution order of the test suite. Such a situation may arise if the regression testing is performed using some form of test selection \cite{LeongSPTM19}.  
A test $t$ is order-dependent if, for the same program and test suite, there exists a passing order $O_p$ for which: 
\[ Run(t,O_p) = Pass\]
And a failing execution order $O_f$ for which:

 \[ Run(t,O_f) = Fail\]
 
The problem of order-dependency has been particularly highlighted when projects updated their Java Development Kit version from $6$ to $7$.
The new Java version modified the reflection implementation and therefore the order adopted by JUnit for running class tests evolved.
Hence, test suites that used to pass correctly, started failing because of the new JUnit orders~\cite{versionIssue1,versionIssue2}.
Accordingly, developers realised that their suites are order-dependent.

As with other categories of flaky tests, order-dependent tests hinder regression testing as they waste development time and introduce uncertainty in the test suite.
Besides these effects, order-dependent tests also impact the effectiveness of techniques that aspire to optimise regression testing.
The study of Lam \etal~\cite{lam2020dependent} showed that order-dependent tests should be accounted for in test selection, prioritisation, and parallelisation techniques.
The dependency between tests makes it difficult to reduce or rearrange the test suite without leading to non-deterministic failures.

\begin{lstlisting}[escapechar=!,label={lst:cleaner}, caption=Example of state polluters and cleaners with a victim test.\\]

//Polluter
$$@Test
public void customConnectionFactory() {
  handler = new RequestHandler() {...};
  ConnectionFactory factory = 
    new ConnectionFactory(){ 
    ... };
  !\colorbox{calmGrey}{HttpRequest.setConnectionFactory(factory);}!
  int code = get("http://not/a/real/url").code();
  assertEquals(200, code);
}

//Cleaner  
$$@Test
public void nullConnectionFactory() {
  handler = new RequestHandler() {...};
  !\colorbox{calmGrey}{HttpRequest.setConnectionFactory(null);}!
  int code = get(url).code();
  assertEquals(200, code);
}

//Victim
$$@Test
public void postWithNumericQueryParams() {
  Map<> inputParams = new HashMap<>();
  inputParams.put(1, 2);
  inputParams.put(3, 4);
  final Map<> outputParams = new HashMap<>();
  handler = new RequestHandler() {...};
 !\colorbox{calmGrey}{HttpRequest request = post(url,inputParams,false);}!
  assertTrue(request.ok());
  assertEquals("2", outputParams.get("1"));
  assertEquals("4", outputParams.get("3"));
}
\end{lstlisting}

Previous studies showed that order-dependency is a result of interactions between different types of tests~\cite{Gyori2015,huo2014improving,Shi2019iFix}.
In the following, we define and explain these types of tests and their role in the passing and failing orders of flaky tests.



\subsection{Polluters, victims, and cleaners}
The study of Gyori \etal~\cite{Gyori2015} identified \textit{polluters} as a source of order-dependent tests.
These tests are not order-dependent themselves but they pollute the state shared between tests and lead to the failure of tests that are executed after them, \aka \textit{victims}.
Accordingly, a \textit{victim} is a test that passes when run alone but fails when run with polluting tests.
Following the formalisation of Shi \etal~\cite{Shi2019iFix}, a test is defined as a victim $v$, if there is a non-empty set of polluters $P$ for which: 
\begin{align*}
    Run(v) = Pass \\
    Run (P,v) = Fail
\end{align*} 

In the same vein, \textit{cleaners} are tests that contain \textit{helper} statements that clean the shared state and allow victims to pass.
A set of tests $C$ are considered cleaners if there exists a victim $v$ and a set of polluters $P$ for which:
\begin{gather*}
    Run (P,C,v) = Pass
\end{gather*}
Listing~\ref{lst:cleaner} shows an example of a test class from the \textit{HTTP Request} project that contains a polluter, a cleaner, and a victim\footnote{\url{https://github.com/kevinsawicki/http-request}}.
In this example, \texttt{customConnectionFactory()} pollutes the shared state by creating a \texttt{ConnectionFactory} and setting it in the static field of the \texttt{HttpRequest}, line 9.
On the other hand, the test \texttt{nullConnectionFactory()} cleans that polluted state by resetting the factory of the \texttt{HttpRequest} to \texttt{null} in line 18.
The test \texttt{postWithNumericQueryParams()} presents a victim because it uses the \texttt{HttpRequest} class without ensuring that its set properly.
As a result, when the victim is run after the polluter, the \texttt{ConnectionFactory} is not null and the assertion in line 32 fails.
However, when the same test is run alone or after the cleaner, the state is properly set and the assertions pass.

\subsection{Brittles}
Huo and Clause~\cite{huo2014improving} identified \textit{brittles} as another form of order-dependent tests.
A \textit{brittle} is a test that fails when run alone but passes when run after other tests.
\textit{Brittles} pass after other tests because they rely on other tests, called \textit{state-setters}, to prepare the state required for their success.
Specifically, a test $b$ is considered as a \textit{brittle} if there is a non-empty set of \textit{state-setters} $S$ that ensures that:
\begin{gather*}
    Run(b) = Fail \\
    Run(S,b) = Pass
\end{gather*}

\section{Execution Plan}
\label{sec:plan}
In this section, we explain our execution plan by first presenting \flaker, our approach for injecting order-dependent tests.
Then, we present the evaluation protocol that we intend to follow in order to answer our research questions.

\subsection{\flaker}
\begin{figure*}[htbp]
\vspace{1.0em}
\centering
\includegraphics[width=\textwidth]{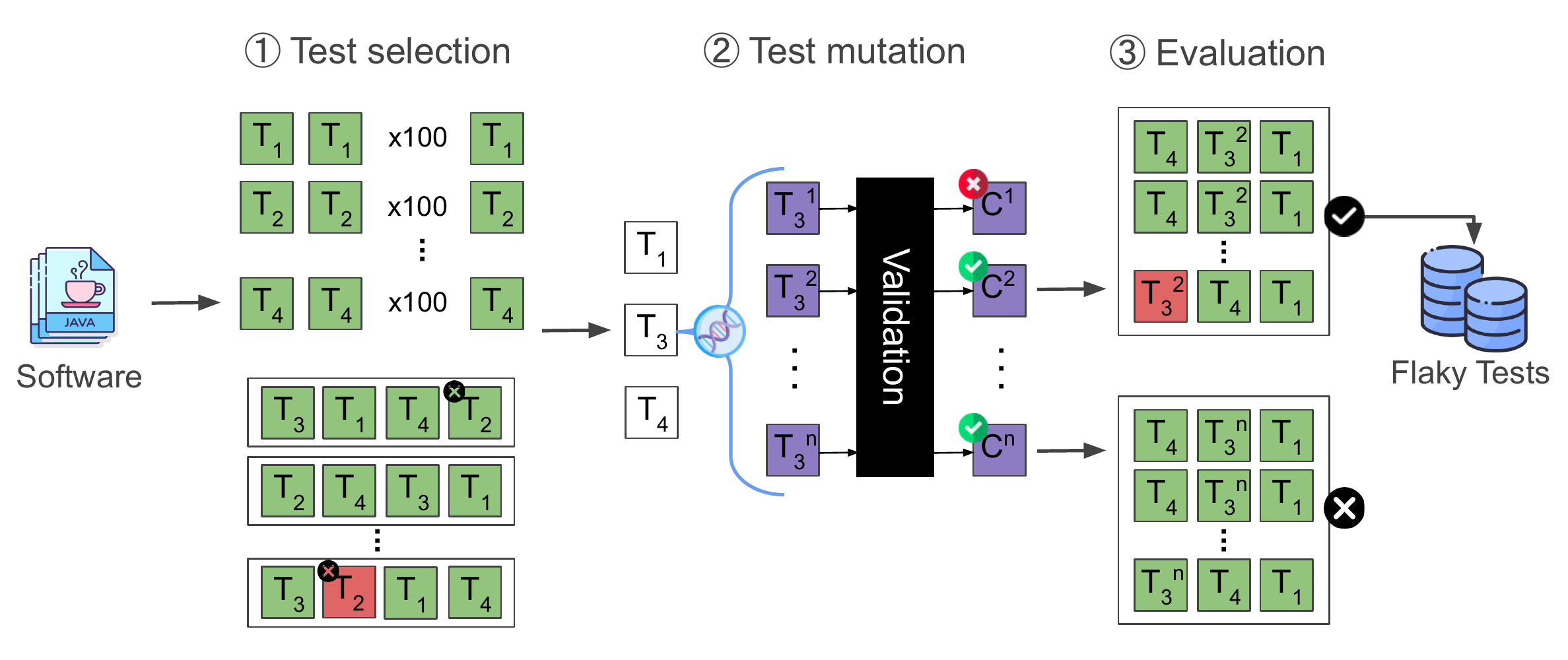}
\caption{Overview of the \flaker approach.}
\label{fig:flaker}
\end{figure*}
Figure~\ref{fig:flaker} presents an overview of the \flaker approach.
\flaker takes as input a Java program with its test suite and outputs a database of order-dependent tests.
First, based on different rerun strategies, \flaker selects stable tests from the program under analysis.
Then, it mutates the selected tests to generate mutants that have potential order-dependency.
Finally, it explores different running orders to evaluate and confirm the order-dependency in the mutant classes.

\subsubsection{\textbf{Step~1:} Select stable tests}
The objective of this step is to filter out tests that are already flaky before proceeding to flakiness injection.
For this purpose, \flaker extracts all tests from the program under analysis and checks their flakiness following a two-step process.

\paragraph{Detect non-order-dependent flaky tests}
Non-order-dependent flaky tests represent tests that are non-deterministic because of general flakiness root causes, \eg concurrency and randomness, and not because of order-dependency.
\flaker identifies these tests by rerunning every test individually $100$ times to check that its outcome is consistent.
If a test result shifts in one of the execution, \ie $Pass \rightarrow Fail$ or $Fail \rightarrow Pass$, the test is considered flaky and we exclude it.

\paragraph{Detect order-dependent tests}
We rely on randomised running orders to detect order-dependent tests.
This strategy allows us to reveal dependencies between tests while avoiding the combinatorial explosion.  
Previous studies showed that randomised orders are the most effective for detecting dependency as they outperform reversal, exhaustive bounded, and dependence-aware bounded algorithms~\cite{zhang2014empirically}.
Since our tool targets Java programs, we only generate orders for tests that are in the same class and we do not consider the whole suite.
We generate $20$ random orders for each test class. 
This choice is based on the experiments of Lam \etal~\cite{Lam2019iDFlakies}, which showed that 20 class orders are sufficient for detecting most flaky tests.



We run each generated order while keeping a record of test outcomes.
Once all orders are run, we analyse the records to spot order-dependent tests, \ie tests that yielded different results over the executions.
We exclude these tests because their non-determinism could influence the conduct of the upcoming steps.

\subsubsection{\textbf{Step 2:} Mutate tests}
The aim of this step is to generate test mutants that could lead to order-dependency.
The study of Shi \etal~\cite{Shi2019iFix} highlighted the presence of \textit{helper} statements that reset the shared state and make tests order-independent.
The objective of \flaker is to remove these \textit{helpers} and make tests order-dependent.
For this, we relied on the mutation operator \textsf{del\_statement}, which allows us to delete any statement from the test code.
With this mutation, we can create mutant tests that are \textit{brittles} or \textit{victims}.


\paragraph{Generate brittles}
Listing~\ref{lst:helper} presents an example of helpers found in the open-source project \textit{Undertow}\footnote{\url{https://github.com/undertow-io/undertow}}.
The example shows that the two tests \texttt{testIdleTimeout()} and \texttt{testCloseReason()} have an identical helper statement in lines 3 and 11 respectively.
These \textit{helpers} reset the state of the \texttt{Endpoint} and ensure that session connections are successful in both tests.

With mutation, we can remove the \textit{helper} in line 11 and make \texttt{testCloseReason()} order-dependent.
Indeed, without resetting the \texttt{Endpoint}, \texttt{testCloseReason()}
would fail when run alone but would pass when run after the \texttt{testIdleTimeout()}.
Consequently, \texttt{testCloseReason()} becomes a \textit{brittle} and \texttt{testIdleTimeout()} becomes a \textit{state-setter}.

\begin{lstlisting}[label={lst:helper}, caption=Example of a helper statement.]
$$@Test
public void testIdleTimeout() throws Exception {
    Endpoint.reset();
    Session session = deployment.connectToServer();
    Assert.assertEquals(..., Endpoint.message());
    ...
}
    
$$@Test
public void testCloseReason() throws Exception {
    Endpoint.reset();
    MessageEndpoint.reset();
    Session session = deployment.connectToServer();
    session.close();
    assertEquals("CLOSED",Endpoint.message());
    ...
}
\end{lstlisting}

\paragraph{Generate victims}

Listing~\ref{lst:polluter} presents a modified example of potential \textit{polluters} and \textit{victims} in the \textit{Hadoop} project\footnote{\url{https://github.com/apache/hadoop/}}.
In this example, the class has a test fixture method that runs once before all class tests.
This method initialises the static variable \texttt{testDir} to prepare for test runs.
On the other hand, \texttt{testWithStringAndConfForBuggyPath()} modifies this variable in line 11 and pollutes the shared state.
However, \texttt{testAbsoluteGlob()} is protected from this pollution thanks to the \textit{helper} in line 19, which reassures that this shared variable is set to the right value before using it in assertions.
Hence, \texttt{testAbsoluteGlob()} is order-independent and it passes even when run after \texttt{testWithStringAndConfForBuggyPath()}.

With mutation, \flaker can remove the helper from \texttt{testAbsoluteGlob()} and transform it into a \textit{victim}.
The test would pass when run alone because the initialisation sets the required state but would fail when run after \texttt{testWithStringAndConfForBuggyPath()}, which becomes a \textit{polluter}.

\begin{lstlisting}[label={lst:polluter}, caption=Example of potential for a polluter and a victim.]


$$@BeforeClass
public static void initialize() {
  ...
  testDir = new Path(System.getProperty(..)+"..");
}

$$@Test
public void testWithStringAndConfForBuggyPath() {
  dirString = "file:///tmp";
  testDir = new Path(dirString);
  assertEquals("file:/tmp", testDir.toString());
  checkPathData();
}

$$@Test 
public void testAbsoluteGlob() {
  testDir = new Path(System.getProperty(..)+"..");
  PathData[] items = PathData.expandAsGlob(...);
  assertEquals(sortedString(
  testDir+"/d1/f1", testDir+"/d1/f1.1"),
  sortedString());
}
\end{lstlisting}

As we cannot identify the \textit{helper} statement beforehand, the mutation is applied to all test statements, except the assertions.
That is, for each test with $m$ non-assert statements, \flaker generates $m$ mutants.
Each mutant test is reinserted in its mother class to form a mutant test class.
If the mutant class does not compile, the mutant is considered invalid and therefore excluded from the analysis.

\subsubsection{\textbf{Step 3:} Evaluate mutants}
The aim of this step is to assess the impact of mutants on order-dependency.
Algorithm~\ref{alg:detection} describes the process of identifying order-dependency in each mutant class.
\begin{algorithm}
\SetAlgoLined
\textbf{Inputs:}
mutantClass\\
\textbf{Outputs:}
flakyTests[]\\
\textbf{Procedure} \emph{identifyFlakiness(mutantClass)}\\

classTests[] = extractClassTests(mutantClass)\\
classOrders[] = generateTestOrders(mutantClass)\\

 \ForEach{ classOrder O $ \in $ classOrders[]}{
   passingTests[], failingTests[] = run(mutantClass, O)
   
   \ForEach{test T $ \in $ classTests[]}{
     \eIf{T $\in$ passingTests[]}
     {T.addPassingOrder(O)}
     { T.addFailingOrder(O)} 
   }
 }
 \ForEach{test T $\in$ classTests[] }{
 \If{ T.passingOrders $\neq \emptyset$ \textbf{and} T.failingOrders $\neq \emptyset$}{
    flakyTests.addTest(T) \\
    \uIf{rerun(T) ==  Pass}{
    T.isVictim = True} 
    \uElseIf{rerun(T) ==  Fail}{
    T.isBrittle = True}
    \Else{
    T.isNonOrderDependent = True
    }
 }
 }
 \textbf{return} flakyTests[]
 \caption{Detect order-dependent tests in mutant.}
 \label{alg:detection}
\end{algorithm}

First, we extract all the tests of the mutant class and we generate random running orders for them.
Then, we run each of these orders and we save the execution results to keep track of orders that led to a test pass or failure.
After executing all the generated orders, we verify if any of the class tests have both passing and failing orders.
If so, this test is flaky and we need to classify it.
For this purpose, we rerun the test 100 times alone to check if it is flaky regardless of the execution order.
In such a case, the test is flaky but not order-dependent and we label it as such.
If the test passes consistently through the reruns, this means that it is a \textit{victim} of other polluting tests, which makes it fail when run with other tests.
However, if the test fails when run alone, we consider it as a \textit{brittle} that needs to run with other \textit{state-setters} to pass.


\subsection{Evaluation}
\begin{table*}[htbp]
\centering
\caption{Evaluation metrics.}\label{table:metrics}
\begin{tabularx}{\textwidth}{lX}
\toprule
\multicolumn{1}{c}{} & \textit{\textbf{Metric}} \\ \midrule

\multirow{8}{*}{{\textbf{RQ1}}} & \cellcolor{calmGrey} \textbf{\#stable tests}: The number of tests selected after filtering out tests that are already flaky in the first step of \flaker. \\
& \textbf{\#mutants}: The total number of mutants generated by \flaker at step 2 (before validation). \\
& \cellcolor{calmGrey} \textbf{\#valid mutants}: The number of mutants that were successfully compiled after their reinsertion in the test class. 
This metric provides insights about the effects of mutation on the test correctness.
\\ 
& \textbf{\#OD mutants}: The number of mutants that show order-dependency at the evaluation step. \\ 

& \cellcolor{calmGrey}\textbf{\%OD tests}: The ratio between the number of order-dependent tests and the total number of tests in a project. \\ 

&\textbf{\%OD classes}: The ratio between the number of order-dependent classes and the total number of test classes in a project. \\ 
& \cellcolor{calmGrey} \textbf{run time}: The average execution time for \flaker on different projects and test classes. \\ \midrule

\multirow{4}{*}{{\textbf{RQ2}}} &  \textbf{\%{brittles}}: The percentage of order-dependent mutants that could not pass if run alone. \\ 
& \cellcolor{calmGrey} \textbf{\%victims}: The percentage of order-dependent mutants that could pass when run alone. \\ 
& \textbf{\%assertion}: The percentage of order-dependent mutants that fail due to assertion errors.\\ 
&  \cellcolor{calmGrey} \textbf{\%exceptions}:The percentage of order-dependent mutants that fail due to other exceptions or errors. \\ \midrule
\multirow{3}{*}{{\textbf{RQ3}}} & \textbf{class size}: The number of tests in a test class.\\ 
&  \cellcolor{calmGrey} \textbf{\#static fields}: The number of static fields in a test class.\\ 
& \textbf{fixture}: A boolean indicating the presence of fixture, \ie \texttt{setUp()} and \texttt{tearDown()} methods, in a test class.\\ 
\bottomrule
\end{tabularx}
\end{table*}

Table~\ref{table:metrics} presents the metrics that we intend to compute after applying \flaker.
In the following, we explain how we leverage these metrics to answer our research questions.
\subsubsection{\textbf{RQ1: How effective is mutation in injecting order-dependent tests?}}
First, we compute the proportion of order-dependent mutants compared to the number of stable tests.
This allows us to assess the capacity of mutation in injecting order-dependent tests.
Also, to highlight the success rate of the mutation, we calculate the proportion of order-dependent mutants compared to all the generated and valid mutants.

To evaluate the suitability of the mutation approach for generating order-dependency datasets, we compare the numbers of generated order-dependent tests to the numbers of order-dependent tests that could be mined using other tools.
Specifically, we compute the metrics \textbf{\%OD tests} and \textbf{\%OD classes} for the order-dependent mutants generated by \flaker.
Then, we compare these values with the proportions of \textbf{\%OD tests} and \textbf{\%OD classes} that could be detected using the tool \textsc{iDFlakies}~\cite{Lam2019iDFlakies}.
These metrics can be retrieved from the results of previous studies, which applied \textsc{iDFlakies} on a set of Java projects.
This comparison allows us to determine whether the mutation approach is a sound alternative for creating order-dependency datasets or not.

Furthermore, to assess the feasibility of the mutation approach, we report the average time consumed by \flaker to perform the mutation and evaluate the order-dependency of mutants.

\subsubsection{\textbf{\textsc{RQ2}: What are the characteristics of injected order-dependent tests?}}
We categorise the order-dependent mutants into \textit{brittles} and \textit{victims} to explain the nature of order-dependent tests that we can inject with mutation.
We also report the causes of failure, \textbf{\%assertion} and \textbf{\%exceptions} to show the impact of mutation on tests.
These four metrics are also useful for the characterisation of the generated dataset and can be useful for studies that rely on it. 
On top of that, these metrics can help in comparing our injected mutants with real order-dependent tests that are observed in the wild.

\subsubsection{\textbf{\textsc{RQ3}: What projects are more receptive to order-dependency injection?}}
We aim to inspect the $14$ projects under study and identify characteristics that could explain their level of proneness to flakiness injection.
For this purpose, we analyse test suite properties, which could be related to order-dependency.
For instance, we compute the size of test classes and analyse its impact on the success of flakiness injection.
We also investigate the effect of shared resources, \eg static fields, which could be a sign of hidden states.
Finally, we inspect the presence of test fixtures, \ie \texttt{setUp()} and \texttt{tearDown()} methods, which could protect tests against our mutation.
To assess the effect of these metrics on the number of injected order-dependent mutants, we use the following statistical tests:
\begin{itemize}[wide=10pt,noitemsep,topsep=0pt]
    \item Spearman's rank coefficient~\cite{zar2005spearman}: a non-parametric measure that assesses how well the relationship between two variables can be described using a monotonic function.
    We use this coefficient to assess the effect of class size and static fields on the number of injected order-dependent mutants.
    This measure does not require the normality of the variables and does not assess the linearity, hence its adequacy for our analysis.
    \item Mann-Whitney U\,test~\cite{sheskin2003handbook}: we use this test to check if the distributions of injected order-dependent tests are identical in the classes that have fixtures or not.
    \item Cliff's\,$\delta$~\cite{romano2006appropriate}: a non-parametric effect size measure, which is suitable for ordinal data.
    We use this measure to quantify the effect size of the presumed difference between the classes based on the presence of fixture.
\end{itemize}
It is worth noting that this question is exploratory and its analysis will depend on the injection results and the project properties.
Hence, the axes presented here are only preliminary and other metrics can be added later.

\section{Dataset}
\label{sec:dataset}
In this section, we present the dataset that we aim to use in our study.

\begin{table}[htbp]
\centering
\caption{Evaluation set.}\label{tab:projects}
\resizebox{\linewidth}{!}{
\begin{tabular}{ll}
\toprule
\textit{\textbf{Project}} & \textit{\textbf{Link}}                                 \\ \midrule
Alien4Cloud               & \url{https://github.com/alien4cloud/alien4cloud}             \\ 
DropWizard                & \url{https://github.com/dropwizard/dropwizard}              \\ 
Dubbo                     & \url{https://github.com/apache/incubator-dubbo}              \\
Elastic-Job-Lite          & \url{https://github.com/elasticjob/elastic-job-lite}        \\ 
Fastjson                  & \url{https://github.com/alibaba/fastjson}                    \\ 
Hsac-Fitness-Fixture      & \url{https://github.com/fhoeben/hsac-fitnesse-fixtures}     \\ 
Junit-quickcheck          & \url{https://github.com/pholser/junit-quickcheck}           \\ 
Marine API                & \url{https://github.com/ktuukkan/marine-api}                \\ 
OpenPOJO                  & \url{https://github.com/OpenPojo/openpojo}                  \\ 
Spring-data-envers        & \url{https://github.com/spring-projects/spring-data-envers} \\ 
Struts                    & \url{https://github.com/apache/struts}                      \\ 
Undertow                  & \url{https://github.com/undertow-io/undertow}               \\ 
Wikidata-Toolkit          & \url{https://github.com/Wikidata/Wikidata-Toolkit}         \\ 
Wildfly                   & \url{https://github.com/wildfly/wildfly}                  \\ \bottomrule
\end{tabular}
}
\end{table}

Table~\ref{tab:projects} shows the projects that we intend to use in our evaluation.
These projects are collected from a public dataset\footnote{\url{https://sites.google.com/view/flakytestdataset}} that was used in previous order-dependency studies.
The original set of projects was first used to identify flaky tests with iDFlakies~\cite{Lam2019iDFlakies}.
Later, Shi \etal used the tool iFixFlakies to patch order-dependent tests in 14 projects of this dataset~\cite{Shi2019iFix}.
In our study, we intend to focus on the same 14 projects to facilitate the early validation of our tooled approach.
Specifically, if our mutation approach is capable of removing \textit{helpers} and evaluating order-dependency accurately, it should be capable of reversing the fixes that were performed by iFixFlakies in these projects.
On top of that, the former existence of order-dependency in these projects could be a favouring factor for our injection.
For instance, these projects can already have shared states or polluting tests but their impact on tests is not visible yet because of the \textit{helpers}.
Our approach could reveal these dependencies by removing \textit{helper} statements.

\section{Threats to Validity}
\label{sec:threats}

A potential threat to our internal validity stems from our strategy for detecting order-dependency.
We opted for randomised orders to detect order-dependent tests in both Step~1 and Step~3 of \flaker.
This strategy is not complete in that it does not explore all possible permutations of tests and it could miss some dependencies.
Thus, it is possible that some order-dependent tests are not filtered at Step~1 and could interfere with our mutant evaluation.
Similarly, some flaky tests could be missed in our mutant evaluation and hinder our study results.
Nevertheless, achieving a complete order-dependency detection requires the exploration of all test permutations, which is impossible due to combinatorial explosion.
The randomised algorithm showed its superiority compared to other incomplete algorithms like order reversal, exhaustive bounded search, and dependence-aware bounded search~\cite{zhang2014empirically}.
Besides, our configuration and choice of the number of reruns align with the positive results observed in previous studies on order-dependency detection~\cite{Lam2019iDFlakies}.

Another possible threat arises from the use of mutation to generate a dataset of flaky tests.
It is possible that the injected tests differ from the ones detected in the wild.
To alleviate this threat, we report several metrics on the injected tests, \eg brittle, victim, and raised exception, to allow the comparison between our injected order-dependent tests and tests from other datasets. 



\section{Conclusion}
\label{sec:conclusion}
This paper presented our planned exploratory study on the injection of flakiness in software tests.
The objective of this study is to evaluate the suitability of test mutation for generating datasets of order-dependent tests. 
To conduct this study, we proposed \flaker, a tooled approach that removes \textit{helper} statements from stable test suites to create order-dependency.
This paper also presented the software projects and analysis protocol that we intend to adopt in our planned study.

\bibliographystyle{IEEEtran}
\bibliography{references}

\end{document}